\documentstyle[11pt,
epsfig]{article}

\begin{document}

\relax

\def\be{\begin{equation}}
\def\ee{\end{equation}}
\def\bs{\begin{subequations}}
\def\es{\end{subequations}}
\def\ut{\tilde{u}}
\def\calm{{\cal M}}
\def\cale{{\cal E}}
\def\lx{\lambda}
\def\ex{\epsilon}
\def\Lx{\Lambda}

\newcommand{\tl}{\tilde t}
\newcommand{\ttt}{\tilde T}
\newcommand{\rhot}{\tilde \rho}
\newcommand{\ptt}{\tilde p}
\newcommand{\drho}{\delta \rho}
\newcommand{\drhot}{\delta {\tilde \rho}}
\newcommand{\dchi}{\delta \chi}
\newcommand{\A}{A}
\newcommand{\B}{B}
\newcommand{\mmu}{\mu}
\newcommand{\mnu}{\nu}
\newcommand{\ii}{i}
\newcommand{\jj}{j}
\newcommand{\jl}{[}
\newcommand{\jr}{]}
\newcommand{\ml}{\sharp}
\newcommand{\mr}{\sharp}

\newcommand{\da}{\dot{a}}
\newcommand{\db}{\dot{b}}
\newcommand{\dn}{\dot{n}}
\newcommand{\dda}{\ddot{a}}
\newcommand{\ddb}{\ddot{b}}
\newcommand{\ddn}{\ddot{n}}
\newcommand{\pa}{a^{\prime}}
\newcommand{\pn}{n^{\prime}}
\newcommand{\ppa}{a^{\prime \prime}}
\newcommand{\ppb}{b^{\prime \prime}}
\newcommand{\ppn}{n^{\prime \prime}}
\newcommand{\fda}{\frac{\da}{a}}
\newcommand{\fdb}{\frac{\db}{b}}
\newcommand{\fdn}{\frac{\dn}{n}}
\newcommand{\fdda}{\frac{\dda}{a}}
\newcommand{\fddb}{\frac{\ddb}{b}}
\newcommand{\fddn}{\frac{\ddn}{n}}
\newcommand{\fpa}{\frac{\pa}{a}}
\newcommand{\fpb}{\frac{\pb}{b}}
\newcommand{\fpn}{\frac{\pn}{n}}
\newcommand{\fppa}{\frac{\ppa}{a}}
\newcommand{\fppb}{\frac{\ppb}{b}}
\newcommand{\fppn}{\frac{\ppn}{n}}

\newcommand{\pt}{\tilde{p}}
\newcommand{\rhb}{\bar{\rho}}
\newcommand{\pb}{\bar{p}}
\newcommand{\pbb}{\bar{\rm p}}
\newcommand{\rht}{\tilde{\rho}}
\newcommand{\kt}{\tilde{k}}
\newcommand{\kb}{\bar{k}}
\newcommand{\wt}{\tilde{w}}

\newcommand{\dA}{\dot{A_0}}
\newcommand{\dB}{\dot{B_0}}
\newcommand{\fdA}{\frac{\dA}{A_0}}
\newcommand{\fdB}{\frac{\dB}{B_0}}

\def\be{\begin{equation}}
\def\ee{\end{equation}}
\def\bs{\begin{subequations}}
\def\es{\end{subequations}}
\newcommand{\een}{\end{subequations}}
\newcommand{\ben}{\begin{subequations}}
\newcommand{\beq}{\begin{eqalignno}}
\newcommand{\eeq}{\end{eqalignno}}

\def \lta {\mathrel{\vcenter
     {\hbox{$<$}\nointerlineskip\hbox{$\sim$}}}}
\def \gta {\mathrel{\vcenter
     {\hbox{$>$}\nointerlineskip\hbox{$\sim$}}}}

\def\rmp{\rm p}
\def\g{\gamma}
\def\mpl{M_{\rm Pl}}
\def\ms{M_{\rm s}}
\def\ls{l_{\rm s}}
\def\l{\lambda}
\def\m{\mu}
\def\n{\nu}
\def\a{\alpha}
\def\b{\beta}
\def\gs{g_{\rm s}}
\def\d{\partial}
\def\co{{\cal O}}
\def\sp{\;\;\;,\;\;\;}
\def\r{\rho}
\def\dr{\dot r}

\def\e{\epsilon}
\newcommand{\NPB}[3]{\emph{ Nucl.~Phys.} \textbf{B#1} (#2) #3}   
\newcommand{\PLB}[3]{\emph{ Phys.~Lett.} \textbf{B#1} (#2) #3}   
\newcommand{\ttbs}{\char'134}        
\newcommand\fverb{\setbox\pippobox=\hbox\bgroup\verb}
\newcommand\fverbdo{\egroup\medskip\noindent%
                        \fbox{\unhbox\pippobox}\ }
\newcommand\fverbit{\egroup\item[\fbox{\unhbox\pippobox}]}
\newbox\pippobox
\def\tr{\tilde\rho}
\def\lb{w}
\def\bbox{\nabla^2}
\def\mt{{\tilde m}}
\def\rct{{\tilde r}_c}

\def \lta {\mathrel{\vcenter
     {\hbox{$<$}\nointerlineskip\hbox{$\sim$}}}}
\def \gta {\mathrel{\vcenter
     {\hbox{$>$}\nointerlineskip\hbox{$\sim$}}}}

\noindent
\begin{flushright}

\end{flushright} 
\vspace{1cm}
\begin{center}
{ \Large \bf The Generalized Dark Radiation \\ and 
Accelerated Expansion in Brane Cosmology
\\} 
\vspace{0.5cm}
{Pantelis S. Apostolopoulos$^{1,2}$ and Nikolaos Tetradis$^2$} 
\\
\vspace{0.5cm}
\it{$^1$Departament de Fisica, Universitat de les Illes Balears,\\ Cra. 
Valldemossa Km 7.5, E-07122 Palma de Mallorca, Spain \\\vspace{0.5cm}
$^2$University of Athens, Department of Physics, \\ Nuclear and 
Particle Physics Section,\\ Panepistimiopolis, Zografos 15771, Athens, Greece.}\\
\vspace{1cm}
\abstract{
The effective Friedmann
equation describing the evolution of the brane Universe in the cosmology
of the Randall-Sundrum model includes a dark (or mirage or Weyl) 
radiation term. 
The brane evolution can be interpreted as the motion of the brane
in an AdS-Schwarzschild bulk geometry. The energy density of
the dark radiation is proportional to the black hole mass.
We generalize this result for an AdS  bulk space with an arbitrary matter
component. We show that the mirage term retains its form, but the black hole
mass is replaced by the covariantly defined integrated mass of 
the bulk matter. As this mass
depends explicitly on the scale factor on the brane, the mirage term
does not scale as pure radiation. 
For low energy densities the brane cosmological evolution 
is that of a four-dimensional Universe with two matter components: the
matter localized on the brane and the mirage matter. There is 
conservation of energy between the two components. This 
behaviour indicates a duality between the bulk theory and a
four-dimensional theory on the brane. 
The equation of state
of the generalized dark radiation is that of a conformal field theory, with
an explicit breaking of the conformal invariance through
the pressure of the bulk fluid.
Accelerated expansion
on the brane is possible only if there is negative pressure on the 
brane or in the bulk, or if the integrated mass of the bulk fluid is
negative. 
\\
\vspace{1cm}
} 
\end{center}


\section{The generalized dark radiation}

In the context of the Randall-Sundrum model \cite{rs},
the Universe is identified with a four-dimensional hypersurface
(a 3-brane) in a five-dimensional bulk with negative cosmological
constant (AdS space). The pertinent action is 
\be
S =\int d^5x\sqrt{-g}\left( \Lambda +M^3R+{\cal{L}}_B\right) +
\int d^4x\sqrt{-\hat{g}}\,\left( -V+{\cal{L}}_b \right),  
\label{action1}
\ee
where $R$ is the curvature scalar of the five-dimensional bulk metric 
$g_{AB}$, $-\Lambda $ the bulk cosmological constant ($\Lambda >0$), $V$ the
brane tension, and $\hat{g}_{\alpha \beta }$ the induced metric on the brane.
(We neglect higher curvature invariants in the bulk and induced gravity
terms on the brane.) The Lagrangian density ${\cal{L}}_B$ describes 
the matter content (particles or fields) 
of the bulk, while the density ${\cal{L}}_b$ describes matter localized on
the brane. 

The geometry is non-trivial (warped)
along the fourth spatial dimension, so that an effective localization of
low-energy gravity takes place near the brane. (No such localization takes
place for the bulk matter.)
For low matter densities on the brane and a pure AdS bulk (no bulk matter), 
the cosmological evolution as 
seen by a brane observer reduces to the standard
Friedmann-Robertson-Walker cosmology 
\cite{binetruy}.
We parametrize the metric as
\begin{equation}
ds^2
=-m^2(\tau,\eta )d\tau^2
+a^2(\tau,\eta )d\Omega_k ^2+d\eta ^2,
\label{gauss}
\end{equation}
with $m(\tau,\eta=0)=1$.
The brane is located at $\eta=0$, while we
identify the half-space $\eta>0$ with the half-space
$\eta<0$. 
The effective Friedmann equation at the location of the brane is
\begin{equation}
H^2=\left(\frac{\dot{R}}{R}\right)^2= \frac{1}{6 M^2_{\rm Pl}}
\left[\tilde{\rho}\left(1+ \frac{\tilde{\rho}}{2V}\right)+\rho_d
\right] -\frac{k}{R^2}+\lambda,  \label{friedmann1}
\end{equation}
with $R(\tau)=a(\tau,\eta=0)$ and $M^2_{\rm Pl}=12M^6/V$. 
The energy density $\rht$ corresponds to matter localized on the brane, 
and arises through the parametrization of the corresponding energy-momentum
tensor 
as $\tilde{T}^A_{~B}=\delta(\eta){\rm diag} (-\rht,\pt,\pt,\pt,0)$.
The contribution $\sim \rht^2$ is typical of brane cosmologies and becomes
negligible for $\rht \ll V$. 
The curvature term ($k=0,\pm 1$)
depends on the geometry of the maximally symmetric
space with constant $\tau$ and $\eta$.
The effective cosmological constant 
$\lambda =(V^2/12M^3-\Lambda )/12M^3$ can be set to zero through an
appropriate fine-tuning of $V$ and $\Lx$.
The conservation equation for the brane energy density is
\be
\dot{\rht}+3H(\rht+\pt)=0.
\label{cons1} \ee

The energy density $\rho_d=(12M^3/\pi^2V)(\calm/R^4)$ depends on
an arbitrary constant $\calm$ and scales as conserved pure radiation.
It is characterized as dark, or mirage, or Weyl radiation 
\cite{binetruy,mirage,hebecker}. Its true nature becomes apparent in 
a Schwarzschild system of coordinates, in which the metric is 
written as
\cite{kraus}
\begin{equation}
ds^2=-n^2(t,r)dt^2+r^2d\Omega _k^2+b^2(t,r)dr^2.  \label{schw}
\end{equation}
In this coordinate system, 
the brane evolution described above corresponds to brane motion 
within a bulk with an
AdS-Schwarzschild geometry, specified by the choice: 
$n^2=1/b^2=\Lx r^2/(12M^3)+k-\calm/(6\pi^2M^3r^2).$  
The constant $\calm$ is identified with the mass of a black hole located
at $r=0$.
It can also be related to the value of the bulk Weyl tensor at the location
of the brane $r=R(\tau)$ (see below). 

This picture can be generalized for an arbitrary bulk energy-momentum
tensor $T^A_{~B}$ using a covariant formalism \cite{covar}.
The details of the calculation are given in ref. \cite{general}.
The brane can be identified with a
4D hypersurface, whose spatial part (denoted by $\cal D$) is 
invariant under a six-dimensional group of 
isometries. 
The spatial curvature is determined by the value of $k$. 
The assumption of maximal symmetry of the spatial part, which 
is essential for our results, implies the existence of a preferred spacelike
direction $e^A$, that represents the local axis of symmetry with respect to
which
all the geometrical, kinematical and dynamical quantities are invariant. 
The preferred spatial direction can be chosen in various ways.
For example, in the Gauss normal coordinate system (\ref{gauss}) 
it is convenient 
to choose the preferred axis of symmetry $\sim \partial_\eta$, while in 
the Schwarzschild system (\ref{schw}) the convenient choice is 
$\sim \partial_r$.
We consider an observer with a 5-velocity $\ut^A$ comoving with the brane. 
The spacelike unit vector field $n^A$ is taken
perpendicular to the brane trajectory ($n_A \ut^A=0$). 
We also consider a bulk observer with a 5-velocity $u^A$ perpendicular to
the preferred direction ($e_A u^A=0$).

Each spatial slice $\cal D$ is covariantly 
characterized by an average length scale
function $\ell$. The derivative of $\ell$ 
along the preferred direction (denoted
by a prime) can
be determined from the relation $D_A e^A \equiv 3\ell'/\ell$, where
$D_A$ is the fully projected, perpendicular to $u^A$, covariant derivative
\cite{general}. In the Schwarzchild system (\ref{schw}), we have $\ell = r$,
while in the Gauss-normal system (\ref{gauss}), $\ell=a(\tau,\eta)$. 
At the location of the brane, $\ell=R(\tau)$. 
The time evolution on the brane can be described in terms of the 
Hubble parameter, defined as $3H=\ut^\alpha_{~;\alpha}$. It can be 
shown \cite{general} that 
\be
H^2=\left(\frac{\dot{\ell}}{\ell}\right)^2
= \frac{1}{6 M^2_{\rm Pl}}
\tilde{\rho}\left(1+ \frac{\tilde{\rho}}{2V}\right) 
+\frac{1}{6\pi^2M^3}\frac{\calm(\ell,\tau)}{\ell^4}
-\frac{k}{\ell^2}+\lambda,  \label{friedmann2}
\end{equation}
where the dot indicates a derivative with respect to the proper time $\tau$
on the brane: $\dot{\ell}=\ell_{;\alpha}\ut^\alpha$.

The quantity $\calm(\ell,\tau)$ is
defined through the relation
\be
\calm=\int_{\ell_0}^\ell 2\pi ^2\rho \ell ^3d\ell 
+ \calm_0, 
\label{mass-function}
\ee
where $\rho\equiv T_{AB}u^Au^B$ is the bulk energy density as measured by 
the bulk observer. We can interpret $\calm(\ell,\tau)$ 
as the generalized comoving
mass of the bulk fluid within a spherical shell
with radii $\ell_0$ and $\ell$. This interpretation is
strictly correct only for $k=1$. However, we shall refer to $\calm$ as 
the integrated mass for all geometries of the spatial slices ${\cal D}$. 
The quantity $\bar{p}\equiv T_{AB}n^An^B$ appearing in eq. (\ref{ray}) is 
the bulk pressure in the direction perpendicular to the brane,
as measured by a brane observer. 
The dependence on the total integrated mass, irrespectively of the specific 
radial dependence of the energy density, is 
reminiscent of the implications of Birkhoff's theorem for 
the gravitational field generated by a matter distribution. 
Both results are consequences of the assumed rotational symmetry of
the geometry, which is inherited by the matter distribution.
If we employ the Schwarzchild system (\ref{schw}), we can set
$\ell_0=r_0=0$. 
Then, the integration constant $\calm_0$ in eq. (\ref{mass-function}) can be
interpreted as the mass of a black hole at $r=0$.

A more intuitive physical interpretation of the contribution $\sim \calm$ in 
eq. (\ref{friedmann2}) can be given for a perfect bulk fluid. In this case
$\calm/(6\pi^2 M^3 \ell^4)=\rho/(12M^3)-\cale/3$, where 
$\cale\equiv C_{ACBD}\ut^An^C\ut^Bn^D$ is a scalar formed out of the 
bulk Weyl tensor \cite{general}. 
Both $\rho$ and $\cale$ are evaluated at the location 
of the brane. It is clear that the bulk affects the brane evolution
through its energy density in the vicinity of the brane. The contribution
$\rho$ is related to the bulk matter, while $\cale$ can be loosely interpreted
as accounting for the effect of the local gravitational field. 
For $\rho=0$ (AdS-Schwarzschild bulk) the whole effect arises through
the gravitational field. This justifies the use of the term Weyl radiation
for the contribution in the effective Friedmann equation.

We return to the general case of an arbitrary bulk energy-momentum
tensor (general fluid).
The sense in which eqs. (\ref{friedmann2}), (\ref{ray})
generalize eqs. (\ref{friedmann1}), (\ref{cons1})
becomes apparent if we rewrite them as
\begin{equation}
H^2=\left(\frac{\dot{\ell}}{\ell}\right)^2= \frac{1}{6 M^2_{\rm Pl}}
\left[\tilde{\rho}\left(1+ \frac{\tilde{\rho}}{2V}\right)+\rho_D
\right] -\frac{k}{\ell^2}+\lambda,  \label{friedmann3}
\end{equation}
\be
\left[\dot{\rht}+3H(\rht+\pt)\right]\left(1+\frac{\rht}{V} \right)=
-\left[\dot{\rho}_D+3H(\rho_D+p_D)\right].
\label{cons3} \ee
(We can set $\ell=R(\tau)$ 
by employing the coordinate system (\ref{gauss}).)
We have defined the effective energy density and pressure of
the generalized dark radiation as
\begin{equation}
\rho _D=\frac{12M^3}{\pi^2V}\frac{\calm}{\ell^4},\hspace{2cm}
p_{D}=\frac{\rho _{D}}{3}+
\frac{8M^3}V\bar{p}.
  \label{effective-matter-pressure}
\end{equation}
For $\rht\ll V$ the system behaves in a remarkably simple fashion.
The brane cosmological evolution 
is that of a four-dimensional Universe with two matter components: the
matter localized on the brane and the mirage matter. There is 
conservation of energy between the two components.

The effective equation of state of the mirage component 
is ${p}_D={p}_D(\rho_D)$. This has a simple form for
$\rho=0$, in which case we recover the 
pure dark radiation with ${p}_D=\rho_D/3$.
In more general cases, the determination of the equation of state requires
the full knowledge of the brane and bulk dynamics.
Similarly, the rate in
which brane matter is transformed into mirage matter cannot be determined
by our considerations. It requires explicit input about the 
interaction between the brane and the bulk matter.
It is also noteworthy that in the general case the equation of state
${p}_D={p}_D(\rho_D)$
depends explicitly on the scale $\ell$. This is a consequence of the
explicit breaking of scale invariance by the bulk distribution, which is
reflected in the theory underlying the mirage component.

\section{Specific models}

Let us consider now some particular examples that confirm the general
picture we presented above. 
Our 
general strategy is to find an explicit solution of the Einstein 
equations in the bulk employing the Schwarzschild coordinates (\ref{schw}),
and then introduce the brane as the bounday of the bulk space.
This can be achieved only if there is an appropriate rate of energy
exchange between the brane and the bulk. As a result, the conservation
of the total energy in the brane and mirage components, as given by
eq. (\ref{cons3}), is guaranteed.
In all the examples, the effective Friedmann equation includes the 
generalized dark radiation term, that has the form of eq. 
(\ref{effective-matter-pressure}) with $\ell=R(\tau)$. 
The function $\calm(R)$ can be determined explicitly 
as the integrated mass $\calm(r)$ in the Schwarzschild frame.
(An additional explicit dependence on time is possible for non-static
bulk geometries, as we shall see below.)

$\bullet$ 
In the simplest example of ref. \cite{bulk1}, the bulk is static
in the Schwarzschild frame and the bulk matter distribution is very
similar to that in the interior of a stellar object. The energy density 
has a profile $\rho(r)$ that can be determined through the 
solution of the Einstein equations. 
The integrated mass $\calm(r)$ of this AdS-star
has a non-trivial dependence on $r$. As a result, the generalized radiation
term $\sim \calm(R)/R^4$ does not scale as radiation. 

$\bullet$ 
In ref. \cite{bulk3} an example of a non-static bulk is given. The bulk
matter is pressureless, but has some initial outgoing velocity in the 
radial direction. The bulk metric is assumed to have the
AdS-Tolman-Bondi form. In Schwarzschild coordinates it is given by 
\begin{equation}
ds^{2}=-dt^2+b^2(t,r)dr^2+S^2(t,r)d\Omega_k^2,
\label{metrictb}
\end{equation}
with $b(r,t)$ given by
\begin{equation}
b^2(t,r)=\frac{S^2_{,r}(t,r)}{k+f(r)},
\label{brttb} \end{equation}
where the subscript denotes differentiation with respect to $r$, and
$f(r)$ is an arbitrary function.
The energy-momentum tensor of the bulk matter has the form
$T^A_{~B}={\rm diag} \left(-\rho(t,r),\, 0,\, 0,\, 0,\, 0  \right).$
The bulk fluid consists of successive shells marked by $r$, whose
local density $\rho$ is time-dependent. 
The function $S(t,r)$ describes the location of the shell marked by $r$
at the time $t$. Notice that
$S(r,t)$ is the actual radial coordinate, while $r$ simply marks the successive
shells. Thus, eq. (\ref{metrictb}) can be put in the form (\ref{schw}), if we
express $r$ as $r=r(S,t)$ and redefine $t$ in order to eliminate the
term $dtdS$ in the metric. It is more convenient, however, to match the
metric (\ref{metrictb}) with (\ref{gauss}) directly, through a
transformation $t=t(\tau,\eta)$, $r=r(\tau,\eta)$ \cite{bulk3}.
 
The Einstein equations reduce to 
\begin{eqnarray}
S^2_{,t}(t,r)&=&\frac{1}{6\pi^2M^3}\frac{\calm (r)}{S^2}-\frac{1}{12M^3}\Lx S^2
+f(r)
\label{tb1} \\
\calm_{,r}(r)&=&2\pi^2 S^3 \rho \, S_{,r}.
\label{tb2} \end{eqnarray}
The integrated mass $\calm(r)$ of the bulk fluid incorporates the
contributions of all shells between 0 and $r$. It can be obtained through
the integration of eq. (\ref{tb2}), in agreement with 
eq. (\ref{mass-function}) for $\ell=S$. 
Because of energy conservation
it is independent of $t$, while $\rho$ and $S$ depend on both $t$ and $r$.
The function $f(r)$ determines the initial radial velocity of the bulk 
fluid, as can be seen from eq. (\ref{tb1}).

It can be shown \cite{bulk3} that 
the effective Friedmann equation for the brane evolution
has the form of eq. (\ref{friedmann3}),
with $\calm=\calm(r(\tau,\eta=0))$ and
$\ell=a(\tau,\eta=0)=S(t(\tau,\eta=0),r(\tau,\eta=0))$.
The mirage term does not scale as pure radiation, but has a complicated
behaviour. The bulk pressure $\bar{p}$ perpendicularly to the brane, as 
measured by a brane observer, obeys $\bar{p} > 0$ \cite{bulk3}, even though
the pressure is zero for a bulk observer comoving with the fluid.
This means that the equation of state of the mirage component has
$p_D > \rho_D/3$. 

$\bullet$ 
Another interesting case is discussed in refs. \cite{bulk2,vaidya}.
The bulk metric is assumed to have the generalized AdS-Vaidya form
\begin{equation}
ds^{2}=-n^{2}(u,r)\, du^{2}+2\ex\,du \,dr+ r^{2}d\Omega_k^2,
\label{metric2}
\end{equation}
where 
\be
n^2(u,r)=\frac{1}{12M^3}\Lx r^2+k-\frac{1}{6\pi^2M^3} \frac{\calm(u,r)}{r^2}. 
\label{ns} \ee
The parameter $\ex$ takes the
values $\ex=\pm 1$.
The energy-momentum tensor of the bulk matter
that satisfies the Einstein equations is
\begin{eqnarray}
{T}^{u}_{~u} = {T}^{r}_{~r} &=& -\frac{1}{2\pi^2} \frac{\calm_{,r}}{r^3}
\label{t002} \\
{T}^{1}_{~1} = {T}^{2}_{~2} = {T}^{3}_{~3} 
&=& -\frac{1}{6\pi^2} \frac{\calm_{,rr}}{r^2}
\label{t112} \\
{T}^{u}_{~r} &=& \frac{1}{2\pi^2} \frac{\calm_{,u}}{r^3},
\label{t042} 
\end{eqnarray} 
where the subscripts indicate derivatives with respect to
$r$ and $u$. The 
various energy conditions are satisfied if 
$\ex \calm_{,u} \geq 0$, $\calm_{,r}\geq 0$,
$\calm_{,rr}\leq 0$, $\calm_{,r}\geq -r\calm_{,rr}/3$.

The preferred axis of symmetry is $\sim \partial_r$, so that the average
length scale function $\ell$ is $\ell=r$. It is then clear from eq. 
(\ref{t002}) that $\calm$ is the integrated mass given by 
eq. (\ref{mass-function}). Another way to reach the same conclusion is to
write the metric in Schwarzschild coordinates 
\begin{equation}
ds^{2}=-n^{2}(t,r)\, dt^{2} +n^{-2}(t,r)\, dr^{2} + r^{2}d\Omega_k^2,
\label{metric3}
\end{equation}
where 
\be
n^2(t,r)=
n^2(u(t,r),r)=
\frac{1}{12M^3}\Lx r^2+k-\frac{1}{6\pi^2M^3} 
\frac{\calm\left(u(t,r),r\right)}{r^2} 
\label{ns2} \ee
and $\partial u/\partial t=1$, $\partial u/\partial r= \ex/n^2$.
The non-zero components of the
energy-momentum tensor that satisfies the Einstein equations for this
metric are given by the same expressions as in eqs. (\ref{t002})-(\ref{t042}),
and the integrated mass is given by eq. (\ref{mass-function}).

It is not surprising, therefore, that the brane evolution is
described by eq. (\ref{friedmann3}) with 
$\calm(\tau,\ell)=\calm(u(\tau,\eta=0),\ell)$.
The effective pressure $p_{D}$, defined in eqs. 
(\ref{effective-matter-pressure}), can be calculated to be
\be
p_D=\frac{4M^3}{\pi^2V}\left(\frac{\calm}{\ell^4}
-\frac{{\partial \calm}/{\partial \ell}}{\ell^3} \right)
-\frac{8M^3\ex}{V}\left(\dot{\rht}+3H(\rht+\pt) \right).
\label{pressvaidya} \ee
For $\rht \ll V$ the second term in the r.h.s. can be neglected. 
The function $\calm(u,r)$ is arbitrary. If it is assumed to have the 
form $\calm=\calm(u)$, the bulk energy-momentum tensor corresponds to
a radiation field. The resulting cosmological solution describes a
brane Universe that exchanges (emits or absorbs) relativistic matter 
with the bulk. For example, 
the form of $\calm(u)$ can be matched to the rate of production
of Kaluza-Klein gravitons during the collisions 
in a thermal bath of brane particles \cite{hebecker,vaidya}. 
In this case, the effective pressure of the mirage component 
becomes $p_D=\rho_D/3$. The system of equations 
(\ref{friedmann3}), (\ref{cons3}) for $\rht \ll V$
describes the evolution of a four-dimensional Universe
with energy exchange between the brane matter component $\rht$ and 
the dark radiation component $\rho_D$.

Other choices for $\calm(u,r)$ that satisfy the energy conditions
are possible as well. If one assumes
$\calm(u,r)=\calm(u)r$,
the mirage component has $p_D=0$ for $\rht \ll V$. 
This is the equation of state of non-relativistic matter.
As a result, 
the mirage component can be characterized as mirage cold dark matter.
However, the physical interpretation of a bulk geometry with 
$\calm(u,r)=\calm(u)r$ remains an open question.
 
$\bullet$ Our final example involves a bulk scalar field in
a global monopole (hedgehog) configuration \cite{bulk3}. 
The field has four components
$\phi^\alpha$, $\alpha=1,2,3,4,$ and its action is invariant under a
global $O(4)$ symmetry.
The field configuration that corresponds to a global monopole is
$\phi^\alpha=\phi(r) x^\alpha/r.$ The asymptotic value 
of $\phi(r)$ for $r\to \infty$
is $\phi_0$.
The metric can be written in Schwarzschild coordinates, as in
eq. (\ref{schw}), with $n=n(r)$, $b=b(r)$ and $k=1$. 
For large $r$, the leading contribution of the monopole
configuration to the energy-momentum tensor comes from the angular
part of the kinetic term. The integrated mass can be calculated to 
be $\calm = 3\pi^2\phi_0^2 \ell^2/2$ for large $\ell$. 
(The global monopole has a diverging
mass in the limit $\ell \to \infty$.)
As a result the effective energy density is 
$\rho_D=18M^3\phi_0^2/(V\ell^2)$.
In this case the mirage component scales $\sim \ell^{-2}$,
similarly to the curvature term. This can be verified by calculating
explicitly the effective pressure $p_D$, which turns out to
be $p_D=-\rho_D/3$ for large $\ell$.

\section{The bulk-brane duality}

We saw in the previous sections that, for low energy densities,
the cosmological evolution on the brane 
is typical of a four-dimensional Universe. 
In addition to the matter localized on the brane,
a mirage matter component appears, which we
characterized as generalized dark radiation. There is conservation of
energy between the two components. 
The nature of the mirage component depends on the bulk matter. In particular,
despite its characterization as generalized dark radiation, the mirage
component can have a very general equation of state. If we define 
$p_D=w_D\rho_D$, there are configurations with
\\
$\bullet$ $w_D > 1/3$: non-relativistic bulk 
matter in an AdS-Tolman-Bondi geometry;
\\
$\bullet$ $w_D = 1/3$: AdS-Schwarzschild bulk geometry; 
AdS-Vaidya bulk with
energy exchage between the brane and a radiation field
in the bulk.
\\
$\bullet$ $w_D = 0$: generalized AdS-Vaidya bulk;
\\
$\bullet$ $w_D =-1/3$: global monopole in an AdS bulk;
\\
$\bullet$ $w_D=-1$: constant field with a non-zero potential in the bulk
(effective cosmological constant).

In the case of an AdS-Schwarzschild bulk the mirage component is pure
radiation. Its appearance can be understood through the AdS/CFT correspondence
\cite{mald,adscft}. 
There is a duality that relates a supergravity theory, 
arising in the low energy limit of an appropriate compactification of
a superstring theory, with a conformal 
field theory.
In particular, the supergravity is defined on the product of a compact
manifold and a five-dimensional manifold $X_5$
with a four-dimensional boundary $M_4$. 
The manifold $X_5$ asymptotically (near the boundary $M_4$) becomes 
an AdS$_5$ space. 
The conformal field theory is defined on $M_4$.
It was suggested in ref. \cite{gubser} that the cosmology in the 
Randall-Sundrum model can be understood through the AdS/CFT correspondence.
The AdS bulk degrees of freedom correspond to a conformal
field theory on the boundary of the bulk space.
The dark radiation term is nothing but the energy density of the conformal
degrees of freedom.

If there are non-zero bulk fields other than the gravitational field, the dual
theory is not expected to be conformal. This is obvious if the
bulk field profile introduces new energy scales, other than the 
fundamental Planck scale $M$ and the cosmological constant $\Lx$.\footnote{
For a pure AdS bulk, we can use the AdS length $L\sim (M^3/\Lx)^{1/2}$ instead
of $\Lx$.} 
As a result, it is not surprising that the effective 
equation of state of the generalized dark radiation can deviate significantly
from that of pure radiation. The remarkable property is that the
brane evolution at low energies 
can be described in four-dimensional terms for any bulk content. 
This implies that the dual description of a bulk gravity theory is
quite general at low energies.

The breaking of conformal invariance is expected to be reflected in the 
trace of the energy-momentum tensor of the generalized dark radiation.
The conformal anomaly gives significant corrections only 
during the early stages of the cosmological evolution, when the
energy density is large. The expected modifications have been
discussed in ref. \cite{elias}.
The explicit breaking of conformal invariance by the bulk matter is apparent
even at low energies. 
According to eq. (\ref{effective-matter-pressure}),
the trace of the energy-momentum tensor is proportional to 
the pressure of the bulk fluid perpendicularly to the
brane as measured by the brane observer.  
In cases in which the duality between a bulk theory with broken conformal
invariance and a boundary theory is known, the trace
can also be expressed through the expectation value
of an operator of the boundary conformal theory. The 
cosmological evolution on the brane at low energies can
be derived either through an explicit solution of the 
five-dimensional Einstein equations or through the 
study of the dual theory in a cosmological context \cite{elias}.

\section{Accelerated expansion in brane cosmology}

An important question concerns the possibility of having accelerated
expansion on the brane as a result of the brane-bulk interaction 
\cite{acceleration1,acceleration2}. More specifically, the inflow of energy from the 
bulk into the brane may induce an evolution similar to that in the steady state
cosmology. In such a scenario, the spontaneous energy creation would
be replaced by the energy inflow from the extra dimension.

In the general framework we are considering, the evolution of $H$ is
determined by the Raychaudhuri equation, that takes the form
\be
\dot{H}=-H^2-\frac{1}{12M^2_{\rm Pl}}
\left[(\rht+3\pt) + \frac{2\rht^2+3\rht\pt}{V}
\right]
-\frac{1}{6\pi^2M^3}\frac{\calm(\ell,\tau)}{\ell^4}
-\frac{1}{M^3} \bar{p}
+\lambda,
\label{ray} \ee
where $\dot{H}=H_{;\alpha}\ut^\alpha$.
The acceleration
parameter is proportional to $\dot{H}+H^2$. 
For this to be positive 
one or more of the following conditions must be satisfied:
\\
a) The effective cosmological constant $\lambda$ is positive.
\\
b) The brane matter satisfies $\tilde{\rho} < V$ and
$\tilde{p}<-\tilde{\rho}/3$.
\\
c) The brane matter satisfies $\tilde{\rho} > V$ and
$\tilde{p}<-2\tilde{\rho}/3$.
\\
d) The integrated mass $\mathcal{M}$ of the bulk matter is negative. 
\\
e) The pressure $\bar{p}$ of the
bulk fluid perpendicularly to the brane, as measured by the
brane observer, is negative.

The first two conditions correspond to the standard 
ways of inducing acceleration in conventional cosmology: a cosmological
constant, or a fluid with sufficiently negative pressure (dark energy). 
The third condition concerns the high-energy regime of brane
cosmology. Negative pressure of the brane matter is again required.
The last two conditions constrain the properties of the bulk matter, as
they are reflected in the generalized dark radiation. It is noteworthy that
the flow of energy towards or from the brane does not appear explicitly
in eq. (\ref{ray}).

The bulk matter affects the cosmological evolution
of the brane in way that is largely independent of its 
detailed distribution.
Inhomogeneneities along the fourth spatial
dimension are integrated out in the definition of the
integrated mass. This mass is then averaged out over the whole
bulk through the combination $\calm/\ell^4$. As a result, 
the cosmological evolution is determined by the average bulk density,
instead of the density in the vicinity of the brane. 
The possibility of a negative integrated mass seems problematic at first
sight, as usually it 
implies the existence of naked singularities or instabilities.
However, counter examples exist in the literature,
such as the negative tension brane in the two-brane model of \cite{rs}.
The correlation between acceleration and negative energy density 
of the dark radiation  
is consistent with the approximate 
cosmological solutions of ref. \cite{acceleration1}. In particular, the
fixed points with accelerated cosmological expansion found 
in ref. \cite{acceleration1}
exist only if the mirage energy density 
is negative. Our analysis shows
that this assumption implies a negative integrated mass for the
bulk matter. 

The last possibility for accelerated expansion requires 
negative bulk pressure perpendicularly to the brane, as measured by 
a brane observer. 
In general, negative pressure is generated with a slowly
evolving, homogeneous field with a potential. 
In such a scenario, the accelerated expansion (or inflation) on the brane
can be induced by a bulk scalar field. 

Of greater interest is the
possibility that a bulk fluid with positive or zero pressure may induce 
accelerated expansion because of its motion towards the 
brane. It can be shown, however, that this is not feasible.  
We have assumed that the spatial part of the brane is invariant under 
a six-dimensional group of isometries. The
most general bulk geometry in
which such a brane can be embedded has a metric that can be put in the form of
eq. (\ref{schw}). There is a bulk observer for 
whom the most general bulk fluid has an energy-momentum
tensor of the form $T^A_{~B}=(-\rho,p,p,p,\rmp)$.
Such an observer is comoving with the bulk fluid, so that the energy fluxes 
are zero.\footnote{The elimination of fluxes is not possible in the 
case of a radiation field. A separate analysis of this case through the 
use of the AdS-Vaidya metric in the bulk shows that accelerated expansion
is not possible without negative pressure \cite{bulk2,bulk3,vaidya}.}
On the other hand, the pressure along the fourth spatial dimension
is not constrained by the assumed symmetries to be equal to the pressure along
the directions parallel to the brane.
It can be easily shown that the pressure $\bar{p}$, measured by the
brane observer, is given by 
\be
\bar{p}=\left(\frac{\partial t}{\partial \eta} \right)^2 \rho
+\left(\frac{\partial r}{\partial \eta} \right)^2 \rmp,
\label{trans} \ee
where $\tau$, $\eta$ are coordinates in the Gauss normal frame of
eq. (\ref{gauss}), and 
the partial derivatives are evaluated at the location of the brane.
It is clear that for $\rho >0$, $\rmp \geq 0$ we always have
$\bar{p} >0$. 
As a result, accelerated cosmological expansion 
on the brane is possible only if negative pressure develops either on
the brane or in the bulk.

\vspace {0.5cm}
\noindent{\bf Acknowledgments}\\
\noindent 
One of us (P.S.A.) gratefully acknowledges the financial support of the 
Spanish ``Ministerio de Educaci\'{o}n y Ciencia'' through research grant No SB2004-0110.
This work was 
supported through 
the research program ``Pythagoras II'' (grant 70-03-7992) 
of the Greek Ministry of National Education.  

\end{document}